\documentclass[useAMS,usenatbib]{mn2e}
\usepackage{graphicx}
\usepackage{rotating}
\usepackage{lscape}
\usepackage{mdwtab}

\newcommand{\kms}{km s$^{-1}$}

\def \arcsec{\hbox{$^{\prime\prime}$}}
\def \arcmin{\hbox{$^{\prime}$}}

\title[The clumpy structure of the L1157 outflow]{The clumpy structure of the chemically active L1157 outflow}
\author[Benedettini et al.]{M. Benedettini$^{1,2}$\thanks{E-mail:
milena@ifsi-roma.inaf.it}, S. Viti$^{2}$, C. Codella$^{3}$, R. Bachiller $^4$, F. Gueth$^5$, M. T. Beltr\'an$^6$, 
\newauthor A. Dutrey$^7$, S. Guilloteau$^7$ \\
$^{1}$INAF-Istituto di Fisica dello Spazio Interplanetario, Area di Ricerca di Tor Vergata, via Fosso del Cavaliere 100, 00133 Roma, Italy \\
$^{2}$Department of Physics and Astronomy, University College London, Gower Street, London, WC1E6BT, UK\\
$^3$INAF-Istituto di Radioastronomia, Sezione di Firenze, Largo E. Fermi 5, 
50125 Firenze, Italy\\
$^4$Observatorio Astron\'omico Nacional (IGN), Apartado 1143, E-28800, Alcal\'a de Henares, Madrid, Spain\\
$^5$Institut de Radio Astronomie Millim\'etrique, 300 Rue de la Piscine, F-38406, Saint Martin d'H\`eres, France\\
$^6$Departament d'Astronomia i Meteorologia, Universitat de Barcelona, Av. Diagonal 647, 08028 Barcelona, Catalunya, Spain\\
$^7$ L3AB, Observatoire de Bordeaux, 2 rue de l'Observatoire, BP 89, 33270 Floirac, France \\}
\begin{document}

\date{Accepted . Received .}

\pagerange{\pageref{firstpage}--\pageref{lastpage}} \pubyear{2006}

\maketitle

\label{firstpage}

\begin{abstract}
We present high spatial resolution maps, obtained with the Plateau de
Bure Interferometer, of the blue lobe of the L1157 outflow. We
observed four lines at 3 mm, namely CH$_3$OH (2$_{\rm K}$--1$_{\rm
K}$), HC$_3$N (11--10), HCN (1--0) and OCS (7--6). Moreover, the
bright B1 clump has also been observed at better spatial resolution in
CS (2--1), CH$_3$OH (2$_1$--1$_1$)A$^-$, and $^{34}$SO
(3$_2$--2$_1$).  These high spatial resolution observations show a
very rich structure in all the tracers, revealing a clumpy structure
of the gas superimposed to an extended emission. In fact, the three clumps detected by previous IRAM-30m single dish observations have been resolved into several sub-clumps and new clumps have been detected in the outflow.  The clumps are
associated with the two cavities created by two shock episodes driven
by the precessing jet. In particular, the clumps nearest the protostar
are located at the wall of the younger cavity with a clear arch-shape
form while the farthest clumps have slightly different observational
characteristics indicating that they are associated to the older shock
episode. The emission of the observed species peaks in different part
of the lobe: the east clumps are brighter in HC$_3$N (11--10), HCN
(1--0) and CS (2--1) while the west clumps are brighter in CH$_3$OH
(2$_{\rm K}$--1$_{\rm K}$), OCS (7--6) and $^{34}$SO
(3$_2$--2$_1$). This peak displacement in the line emission suggests a
variation of the physical conditions and/or the chemical composition
along the lobe of the outflow at small scale, likely related to the shock
activity and the precession of the outflow. In particular, we observe the decoupling of the silicon monoxide and methanol emission, common shock tracers, in the B1 clump located at the apex of the bow shock produced by the second shock episode.
\end{abstract}


\section{Introduction}

L1157-mm is a Class 0 protostar located at 440 pc, with a mass of $\sim$
0.12 $M_{\rm \odot}$ (\citealt{gueth97}; \citealt{beltran04}), driving
a spectacular bipolar outflow. The outflow has been studied in detail
through many molecular lines, such as CO (\citealt{umemoto92};
\citealt{gueth96}), SiO (\citealt{zhang95}, 2000; \citealt{gueth98};
\citealt{bachiller01}; \citealt{nisini07}), H$_2$ (\citealt{hodapp94};
\citealt{davis95}), NH$_3$ \citep{tafalla95}, and CH$_3$OH
(\citealt{avery96}; \citealt{bachiller01}).

With respect to other outflows driven by low-mass protostars, L1157
stands out for its rich mm-wave spectrum and it can be considered as
the prototype of chemically active outflows. \cite{bachiller97} have
mapped the L1157 outflow in many molecular mm-emission lines and six
low-velocity molecular clumps have been detected along the
lobes. These clumps have strong emission due to molecular species
which increase their abundances only when a shock-induced chemistry is
active \citep{bachiller01}. Two main shock events have been detected
in the blue lobe of the outflow and the interferometric image of the
CO (1--0) line \citep{gueth96} reveals the presence of two prominent
limb-brightened cavities created by the propagation of large
bow-shocks. The different orientation of the two cavities testifies
the precession of the driving, highly-collimated jet.

Clumpiness in chemically active outflows has been studied recently
(\citealt{viti04}; \citealt{benedettini06}) by fitting the line
profiles of several species by the use of a chemical and a radiative
transfer model. These studies indicate that the size derived from
single dish observations ($\sim$ 0.12 pc for CB3 and $\sim$ 0.04 pc for
L1157) are upper limits and that the clumps contain substructures. In
absence of observational evidence of the structure of the low velocity
clumps in outflow, \cite{viti04} and \cite{benedettini06} assumed the
density profile derived for starless cores in their
modelling. However, the outflow clumps show fundamental differences
with respect to starless cores, for example in the temperature
(starless cores have temperature $\le$ 10 K) and in the chemical
structure (in starless cores most of the molecular material is
depleted toward the center of the core). Direct evidence of the real
structure of these low velocity clumps can only be obtained via very
high spatial resolution observations using an interferometer.

In order to investigate the small scale structure of the low velocity clumps in outflows we carried out interferometric observations of the nearby L1157 outflow with the IRAM Plateau de Bure Interferometer (PdBI) and the results are presented in this paper. We mapped the blue lobe, which is the richest in terms of molecular emission, in a sub--sample of the species already observed with the IRAM-30m single dish in this outflow. In particular, we selected molecules that are high density tracers (as CS, HCN and HC$_3$N) and whose chemical abundance is significantly affected by the presence of a no--dissociative shock (as CH$_3$OH and OCS). 
The observations are described in Sect. 2. We present the results and the calculation of the column densities in Sect. 3 and in Sect. 4, respectively. The results are discussed in Sect. 5 and the main conclusions of the paper are summarized in Sect. 6.

\begin{table*}
  \caption{List of the observed transitions and observing parameters. The first three lines were observed only in the central part of the blue lobe, the B1 clump.}
  \begin{tabular}{@{}lrccccc@{}}
  \hline
 transition & $\nu$ & resolution & PdBI          & clean beam & rms & short-spacings\\
            & GHz   & \kms       &configuration  & arcsec     & Jy beam$^{-1}$ \kms &\\
 \hline
CH$_3$OH (2$_1$--1$_1$)A$^-$ & 97.583 & 0.5 & 5D+6Cp & 3.14 $\times$ 2.82 (P.A.=69\degr) & 0.003 & N\\
$^{34}$SO (3$_2$--2$_1$)& 97.715 & 0.5 & 5D+6Cp & 3.14 $\times$ 2.81 (P.A.=69\degr) & 0.003 & N\\
CS (2--1)               & 97.981 & 0.5 & 5D+6Cp & 3.11 $\times$ 2.79 (P.A.=70\degr) & 0.009 &N \\
\hline
OCS (7--6)             & 85.139 & 0.28 & 5C2+4D1+5D & 6.68 $\times$ 5.50  (P.A.=-79\degr)& 0.05 & N\\
HCN (1--0)             & 88.632 & 0.27 & 5C2+4D1+5D & 5.63 $\times$ 4.28 (P.A.=93\degr)  &0.80 & Y \\
CH$_3$OH (2$_{\rm K}$--1$_{\rm K}$)& 96.741$^{\dag}$ & 0.24 & 5C2+4D1+5D & 5.06 $\times$ 4.37 (P.A.=-85\degr) & 1.30 & Y \\
HC$_3$N (11--10)       & 100.076& 0.23 & 5C2+4D1+5D & 3.84 $\times$ 3.51 (P.A.=-43\degr) &0.14 & N\\
\hline
\end{tabular}\\
\label{obs}
$^{\dag}$ frequency of the (2$_0$--1$_0$)A$^+$ transition
\end{table*}

\section{Observations}

The interferometric observations were carried out with the IRAM
interferometer at Plateau de Bure in three different periods and
configurations. The first set of observations was carried out between
May and July 1997, the second between December 1997 and March 1998 and
the third between September and November 2004. Some preliminary results from the first and second data sets were presented by \cite{Perez-Gutierrez99}, but we report here final images together with a detailed analysis of the three data sets. In the first two periods the CD configuration was used with three configurations of 4 or 5 antennas (5C2+4D1+5D) and a baseline between 24 m and 176 m, with a final clean beam of 6.7$\times$5.5 arcsec$^2$ at 85 GHz and 3.8$\times$3.5 arcsec$^2$ at 100 GHz. A mosaic of 5 fields was performed to cover all the south lobe of the L1157 outflow at 3 mm. The correlator units were configured in order to have a resolution of 78 kHz, which corresponds to 0.25 \kms\, at 3 mm.
In a first setup we observed CH$_3$OH (2$_{\rm K}$--1$_{\rm K}$) at 96.741 GHz in the LSB and HC$_3$N (11--10) at 100.076 GHz in the USB and in a second observation HCN (1--0) at 88.632 GHz and OCS (7--6) at 85.139 GHz were observed simultaneously. The CH$_3$OH (5$_{\rm K}$-4${\rm _K}$) line at 241.285 GHz was observed simultaneously. The phase and amplitude calibration was achieved by observations of 2021+614, which is close to L1157 in the sky. The band pass of the receivers were calibrated by observations of 3C273 and 3C454.3. To correct for decorrelation due to phase noise, amplitude calibration was also done relative to 2021+614, whose flux was determined relative to 3C273 and 3C454.3.  In the observations carried out in 2004 the CD configuration was used with 6 or 5 antennas (5D+6Cp) obtaining a better spatial resolution of $\sim$ 3.1$\times$2.8 arcsec$^2$ at 97 GHz. One pointing on the central zone of the blue lobe, the B1 clump [R.A.(J2000) $20^{\rm h} 39^{\rm m} 9\fs5$; dec(J2000) 68$\degr$01$\arcmin$15$\arcsec$], was performed to observe simultaneously CS (2--1) at 97.981 GHz, CS (5-4) at 244.935 GHz and CH$_3$OH (5$_{\rm K}$-4${\rm _K}$) at 241.791 GHz. The $^{34}$SO (3$_2$--2$_1$) at 97.715 GHz and CH$_3$OH (2$_1$--1$_1$)A$^-$ at 97.583 GHz were automatically covered by the setting of the autocorrelator units that were configured in order to have a resolution of 0.5 \kms\, at 3 mm. The phase and amplitude calibration was achieved by observations of 2037+511 and 1928+738, which are close to L1157 in the sky. The band pass of the receivers was calibrated by observations of 2145+067 and 1928+728. The flux calibration was determined relative to MWC349. The data were calibrated and analyzed with the GILDAS software, the images were produced using natural weighting and cleaned in the usual way. The details of the observations are summarized in Table \ref{obs}. 
The 1.3 mm data (not shown) have a very low signal to noise ratio (S/N), this does not allow us to properly analyse those data.

In the interferometers the emission from structures more extended than 2/3 of the primary beam is filtered out. The extended emission, that is expected to be present in outflows, could be partially recovered by adding single dish data to the interferometric observations. To this aim, we include the short-spacings to the PdBI data for the two lines CH$_3$OH (2$_{\rm K}$--1$_{\rm K}$) and HCN (1--0), for which single dish IRAM-30m data are available \citep{bachiller01}.

\begin{figure*}
\includegraphics[width=6.3cm,angle=-90]{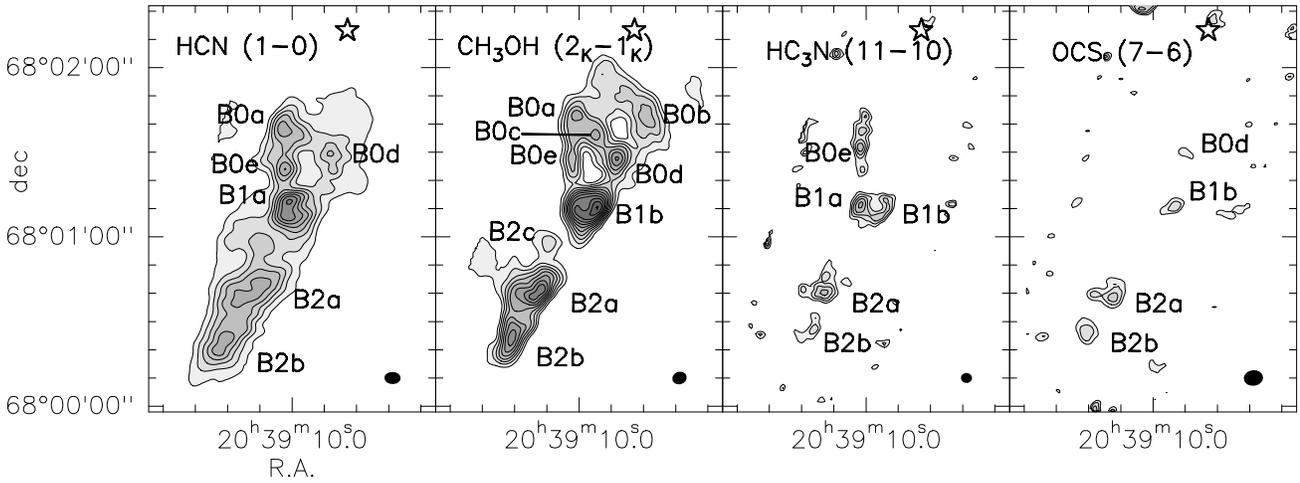}
\caption{Interferometric images of the blue lobe of the L1157 outflow. Note that for HCN (1--0) and CH$_3$OH (2$_{\rm K}$--1$_{\rm K}$) the single dish data have been added to the interferometric data. The star indicates the position of the millimeter driving source and the ellipse at the bottom right shows the clean beam. The contours are: for HCN (1--0) first contour 2.5 Jy beam$^{-1}$ \kms\, (3$\sigma$), level steps 2 Jy beam$^{-1}$ \kms;  for CH$_3$OH (2$_{\rm K}$--1$_{\rm K}$) first contour 4 Jy beam$^{-1}$ \kms\, (3$\sigma$), level steps 1.5 Jy beam$^{-1}$ \kms; for HC$_3$N (11--10) first contour 0.4 Jy beam$^{-1}$ \kms\, (3$\sigma$), level steps 0.15 Jy beam$^{-1}$ \kms\, and for OCS (7--6) first contour 0.15 Jy beam$^{-1}$ \kms\, (3$\sigma$), level steps 0.05 Jy beam$^{-1}$ \kms.}
\label{all_lobe}
\end{figure*}

\begin{figure*}
\includegraphics[width=6cm,angle=-90]{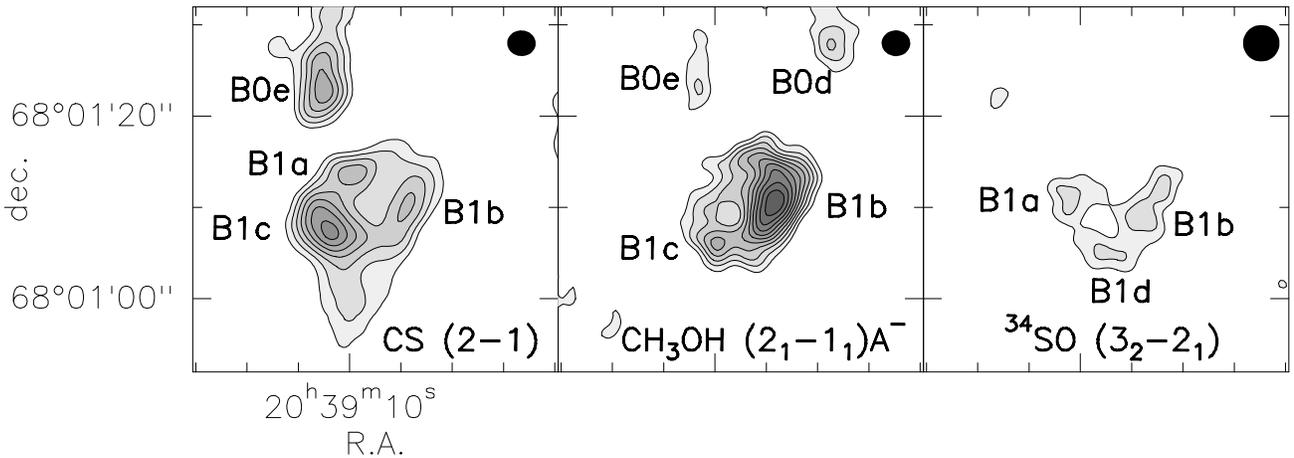}
\caption{Maps of the B1 clump of the L1157 outflow. The contours are: for CS (2--1), first level is 0.027 Jy beam$^{-1}$ \kms (3$\sigma$), level steps are 0.02 Jy beam$^{-1}$ \kms; for CH$_3$OH (2$_1$--1$_1$)A$^-$, first level is 0.009 Jy beam$^{-1}$ \kms (3$\sigma$), level steps are 0.005 Jy beam$^{-1}$ \kms; for $^{34}$SO (3$_2$--2$_1$), first level is 0.01 Jy beam$^{-1}$ \kms (3$\sigma$), level steps are 0.004 Jy beam$^{-1}$ \kms. The ellipse at the top right shows the clean beam. The $^{34}$SO map has been convolved at a higher spatial resolution with respect to the observations, in order to improve the S/N.}
\label{b1}
\end{figure*}

\section{Results}

\subsection{Overall structure}

The maps of the blue lobe of the L1157 outflow in HCN (1--0), HC$_3$N (11--10), CH$_3$OH (2$_{\rm K}$--1$_{\rm K}$) and OCS (7--6) are shown in Fig. \ref{all_lobe}. The central part of the outflow has also been mapped in three more lines, namely CS (2--1), CH$_3$OH (2$_1$--1$_1$)A$^-$ and $^{34}$SO (3$_2$--2$_1$) whose maps are shown in Fig. \ref{b1}. These high spatial resolution observations reveal the presence of a clumpy structure superimposed to an extended emission. 

The extended emission is clearly visible in the two maps, HCN (1--0) and CH$_3$OH (2$_{\rm K}$--1$_{\rm K}$), where the short-spacings are included, while it is not present in the other maps, where it is probably filtered out by the interferometer. The extended emission matches very well with previous observations, namely multi-species single dish maps \citep{bachiller01} and interferometric CO (1--0) map \citep{gueth96}, tracing the large-scale outflow structure.

The interferometric observations show also a very rich structure at small-scale, revealing a clumpiness of the molecular gas richer than the one previously observed.
The three large molecular clumps, called B0, B1 and B2, previously detected by single dish observations \citep{bachiller01} have been resolved into more sub-clumps: the B0 clump is resolved into three sub-clumps labelled B0a, B0c and B0e, the B1 clump is resolved into four sub-clumps labelled B1a, B1b, B1c and B1d, and the B2 clump is resolved into two sub-clumps labelled B2a and B2b. Moreover, three new clumps have been detected: B0b at north-west, B0d located between B0b and B1 and B2c located between B1 and B2. The coordinates of the identified clumps are listed in Table \ref{clumps_coor}; they have been defined by using the tracer where the clump is best defined.
Some of these clumps are associated with the shock front where the outflow jet impacts the ambient medium while others are in the wake of the shocked region where the matter has already been affected by the passage of the shock (see Sect. \ref{disussion_clumps} for a detailed description).

\begin{table}
\caption{Coordinates of the clumps. In the last column the transition used to define the coordinates is listed.}
\begin{tabular}{@{}lccc@{}}
 \hline
clump & R.A.(J2000) & Dec(J2000) & transition\\
    & h~ m ~s~~~  & $\degr$ ~ $\arcmin$~~ $\arcsec$\\
 \hline\\
B0a & 20 39 10.4 & 68 01 38 & HCN (1--0)\\
B0b & 20 39 05.7 & 68 01 42 & CH$_3$OH (2$_{\rm K}$--1$_{\rm K}$)\\
B0c & 20 39 09.0 & 68 01 36 & CH$_3$OH (2$_{\rm K}$--1$_{\rm K}$)\\
B0d & 20 39 07.7 & 68 01 28 & CH$_3$OH (2$_{\rm K}$--1$_{\rm K}$)\\
B0e & 20 39 10.3 & 68 01 24 & CH$_3$OH (2$_{\rm K}$--1$_{\rm K}$)\\
B1a & 20 39 10.2 & 68 01 12 & HCN (1--0)\\
B1b & 20 39 08.8 & 68 01 10 & CH$_3$OH (2$_{\rm K}$--1$_{\rm K}$)\\
B1c & 20 39 10.4 & 68 01 07 & CS (2--1)\\
B1d & 20 39 09.2 & 68 01 04 & $^{34}$SO (3$_2$--$2_1$)\\
B2c & 20 39 11.8 & 68 00 58 & CH$_3$OH (2$_{\rm K}$--1$_{\rm K}$)\\
B2a & 20 39 12.6 & 68 00 40 & CH$_3$OH (2$_{\rm K}$--1$_{\rm K}$)\\
B2b & 20 39 14.1 & 68 00 24 & CH$_3$OH (2$_{\rm K}$--1$_{\rm K}$)\\
\hline
\end{tabular}
\label{clumps_coor}
\end{table}

\begin{figure*}
\includegraphics[width=10.4cm,angle=-90]{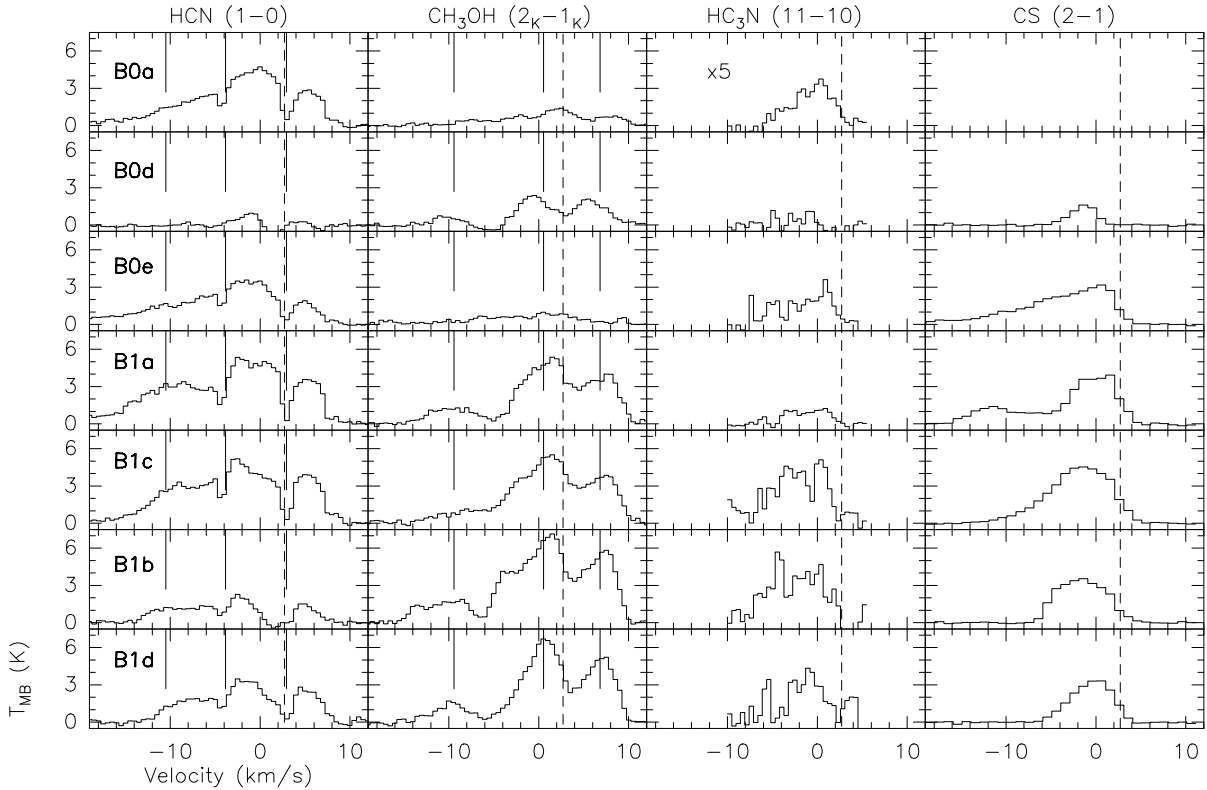}
\caption{Spectra of the HCN (1--0), CH$_3$OH ((2$_{\rm K}$--1$_{\rm K}$), HC$_3$N (11--10) and CS (2--1) in the B0a, B0d, B0e, B1a, B1c, B1b, and B1d clumps. The $T_{\rm MB}$ of the HC$_3$N line is multiplied by a factor of 5. The dashed line is the $v_{\rm lsr}$=2.7 \kms. The continuous lines mark the position of the three hyperfine transitions of HCN (1--0) and the three transitions CH$_3$OH (2$_{-1}$--1$_{-1}$)E, (2$_0$--1$_0$)A$^+$ and (2$_{0}$--1$_{0}$)E.}
\label{spettri}
\end{figure*}

In Fig. \ref{spettri} we show the spectra of HCN (1--0), CH$_3$OH (2$_{\rm K}$--1$_{\rm K}$), HC$_3$N (11--10) and CS (2--1) in some of the identified clumps. At the spectral resolution of the observations the three lines of the hyperfine structure of the HCN (1--0) transition at 88.630 GHz, 88.632 GHz and 88.634 GHz are resolved. The three lines are strongly self-absorbed along the flow but not at the position of the mm source. The velocity of the self-absorption features remains constant along the lobe suggesting the presence of an extended ambient component of HCN; alternatively the self-absorption might be due to high opacity effects. In the spectra of the CH$_3$OH (2$_{\rm K}$--1$_{\rm K}$) transition (see Fig. \ref{spettri}) three lines are detected: (2$_{-1}$--1$_{-1}$)E at 96.739 GHz,
(2$_0$--1$_0$)A$^+$ at 96.741 GHz and (2$_{0}$--1$_{0}$)E at 96.745 GHz. The three lines are blended in most of the map.

The clumps are not detected in all the species. In fact, the S/N of the maps is not the same due to the fainter signal of some lines. In particular, the maximum S/N in the map is higher for HCN (1--0), CH$_3$OH (2$_{\rm K}$--1$_{\rm K}$) and CS (2--1) (19, 18 and 16 respectively) while it is lower for HC$_3$N (11--10), OCS (7--6) and $^{34}$SO (3$_2$--2$_1$) (7, 5 and 4 respectively).

The CH$_3$OH (2$_{\rm K}$--1$_{\rm K}$) and HCN (1--0) emission shows the presence of an extended molecular component (note that these maps include the short-spacings) and they also have the richest structure in term of clumps (see Fig. \ref{all_lobe}). In particular, all the identified clumps, but B0b and B0c, are detected in these tracers. B0b and B0c are detected only in the methanol lines. 
The HC$_3$N (11--10) emission has a definitely lower S/N; nevertheless 5 clumps are clearly identified (see Fig. \ref{all_lobe}). In the northern region of the map the peak of the HC$_3$N (11--10) emission is at an intermediate position between the B0a and B0e clumps identified by HCN. The B1a clump is brighter with respect to the B1b. The two clumps B2a and B2b have also been detected in HC$_3$N; however, the B2b peak is shifted toward the north-west with respect to the peak of HCN (1--0) and CH$_3$OH (2$_{\rm K}$--1$_{\rm K}$).
Despite the low S/N of the OCS map (see Fig. \ref{all_lobe}), the two most southern clumps B2a and B2b are clearly identified, while only a weak signal can be seen in B0d  and B1b.  The quality of the $^{34}$SO map (see Fig. \ref{b1}) is also poor, however a clear arch-shape emission is detected. Three sub-clumps are detected above the noise: B1a and B1b have also been seen in other tracers, while a new clump, B1d is detected at an intermediate position between B1b and B1c. This clump is clearly defined only in this tracer, however, given the low S/N of the peak, we cannot rule out the possibility that it is an artifact. Despite the low S/N, it seems that hints of $^{34}$SO and OCS emission are mainly peaking towards the western part of the B1 clump and in the southern part of the blue lobe.

\begin{figure*}
\includegraphics[width=13cm,angle=-90]{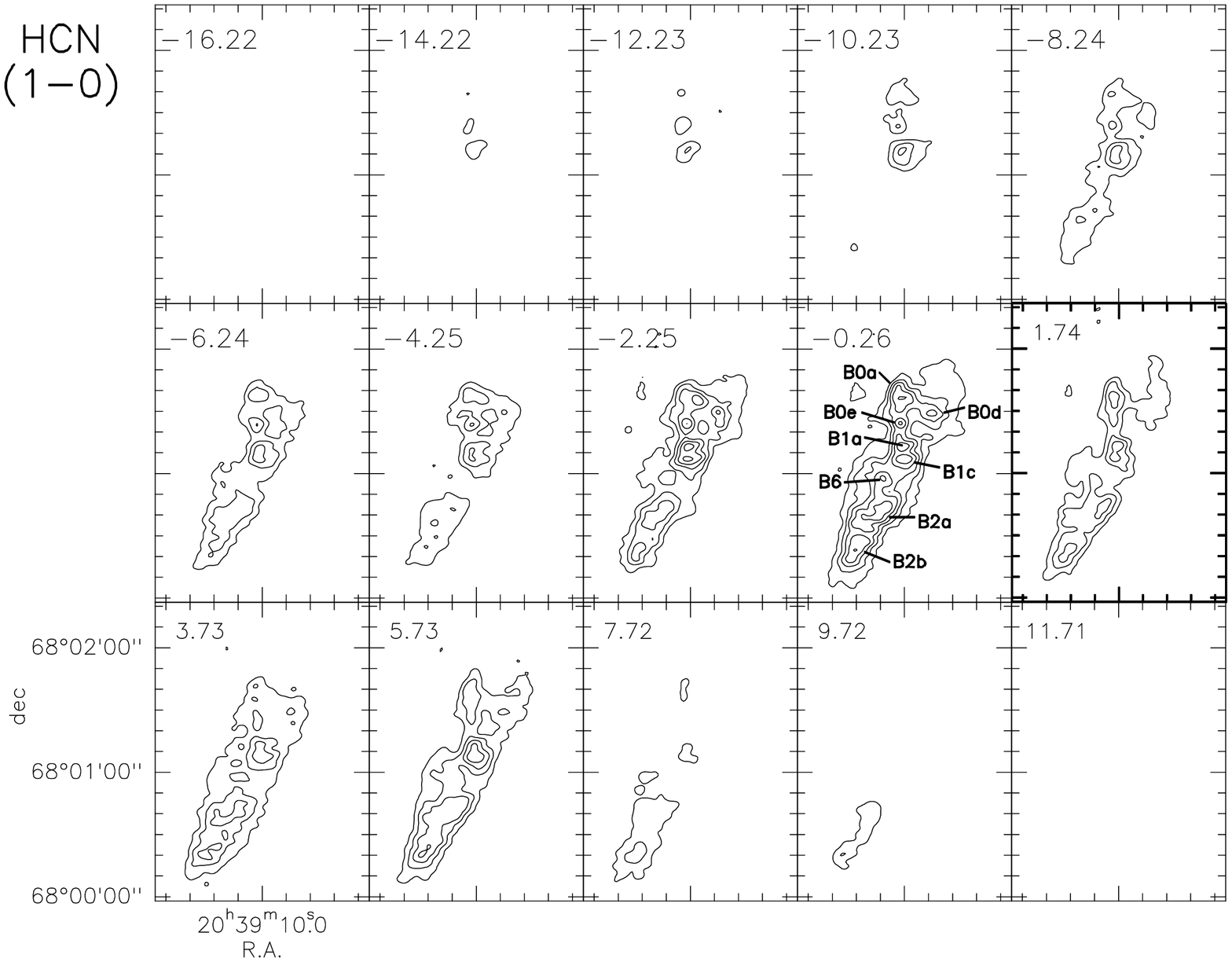}
\caption{Channel map of the blue lobe of the L1157 outflow in the HCN (1--0) line at 88.632 GHz. The thick box points out the ambient velocity emission. First level and level steps are 0.2 Jy beam$^{-1}$ (3$\sigma$).}
\label{hcn_can}
\end{figure*}

Finally, note that most of the clumps are quasi-circular with size of 10 -- 15 arcsec (0.02 -- 0.03 pc at a distance of 440 pc), apart from B2a and B2b that are elongated with size of $\sim$ 10$\times$20 arcsec$^2$. The emission in the most southern clumps is at higher velocity with respect to the northern clumps. This trend can be seen in the channel map of HCN in Fig. \ref{hcn_can} and it is also present in all the other species (channel maps not shown).

\subsection{The clump B0}

B0a, B0b, B0c, B0d  and B0e are the clumps nearest to the protostar driving the outflow. B0a can be identified with the peak of the B0 clump observed in the single dish observations. Close to B0a we detect other clumps: B0c and B0d at south-west of B0a, B0e at south, and B0b at west. B0b and B0c are detected only in CH$_3$OH while B0a is detected also in the HCN (1--0) (note that these clumps are not mapped in CS and $^{34}$SO). B0d is detected in HCN (1--0) and CH$_3$OH and emits marginally in OCS (7--6) and CS (2--1), while B0e is detected in all the tracers but OCS and $^{34}$SO. The HC$_3$N (11--10) line shows an elongated emission covering B0a and B0e with a peak at an intermediate position between the two. 

\subsection{The clump B1}

\begin{figure*}
\includegraphics[width=12cm,angle=-90]{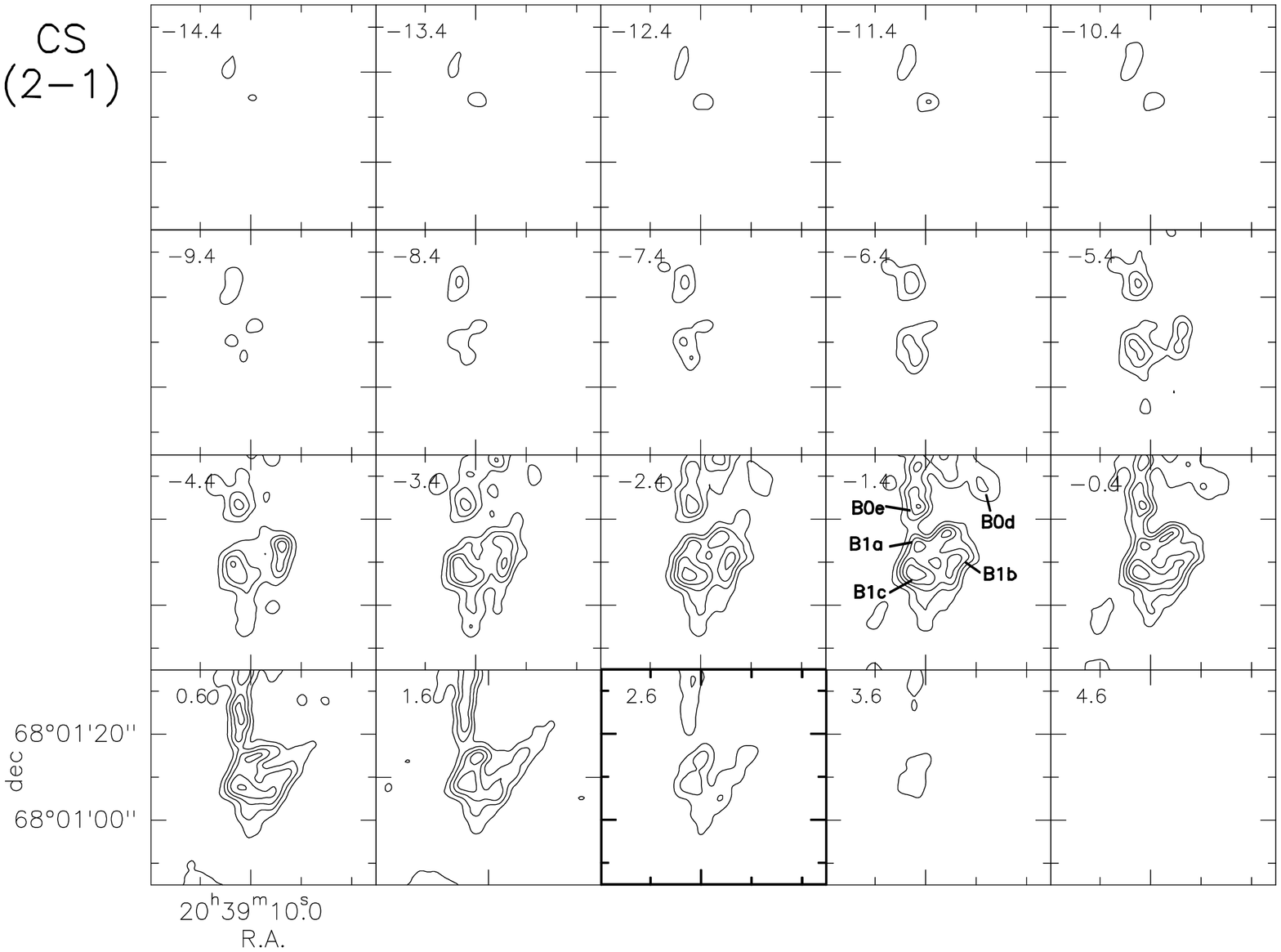}
\caption{Channel map of the B1 clump of the L1157 outflow in the CS (2--1) line at 97.98 GHz. The thick box points out the ambient velocity emission. First level is 0.05 Jy beam$^{-1}$ (3$\sigma$), level steps are 0.07 Jy beam$^{-1}$.}
\label{cs_can}
\end{figure*}

\begin{figure*}
\includegraphics[width=5.7cm,angle=-90]{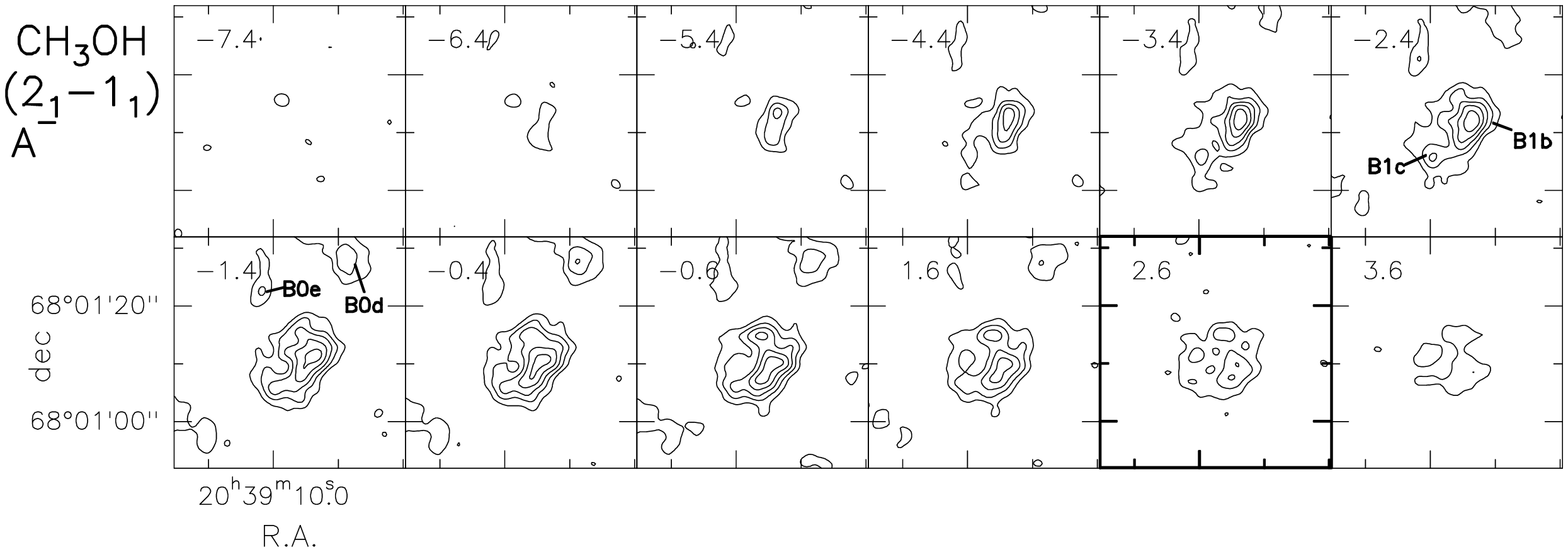}
\caption{Channel map of the B1 clump of the L1157 outflow in the CH$_3$OH (2$_1$--1$_1$)A$^-$ at 96.582 GHz. The thick box points out the ambient velocity emission. First level is 0.01 Jy beam$^{-1}$  (3$\sigma$), level steps are 0.015 Jy beam$^{-1}$.}
\label{ch3oh3mm_can}
\end{figure*}

B1 is the brightest clump of the outflow and it has been mapped in CH$_3$OH (2$_1$--1$_1$)A$^-$, CS (2--1) and $^{34}$SO (3$_2$--2$_1$) with a spatial resolution higher than the other maps. B1 has a well defined arch-like shape with four sub-clumps (B1a, B1c, B1d and B1b) outlining the arch. The CH$_3$OH (2$_1$--1$_1$)A$^-$ emission shows the same quasi-circular morphology of the larger CH$_3$OH (2$_{\rm K}$--1$_{\rm K}$) map with a peak in B1b while the CS emission shows three clumps located in an arched-like shape with B1c showing the strongest emission. The arch shape emission is also detected in the HC$_3$N (11--10) and $^{34}$SO (3$_2$--2$_1$) lines.
In CS (2--1) we detect a diffuse component that is elongated in the south direction, similarly to what is seen in the SiO map by \cite{gueth98}. The elongated feature is clearly visible in the channels map (Fig. \ref{cs_can}) for velocities higher than $-5$ \kms. The channel map also shows that close to the systemic velocity ($v_{\rm lsr}$ = 2.7 \kms), the diffuse emission assumes a more pronounced arch-like shape and that the B0e clump becomes elongated toward B1a and then it `connects' to the B1 clump at $v$ = $-1.4$ \kms.  In the B1a position a second CS (2--1) velocity component is identified in addition to the main component centered at 0 \kms (see Fig. \ref{spettri}); the second component peaks at $-11.4$ \kms\, and extends up to $-16$ \kms. This component is also present in the SiO (3--2) line observed by \cite{bachiller01}. In the channel map of the CH$_3$OH (2$_1$--1$_1$)A$^-$ line (shown in Fig. \ref{ch3oh3mm_can}) one can see that the peak position of the B1b clump is shifted toward the south at higher velocity. Indeed, the emission peaks at an offset of ($-4\arcsec$,$4\arcsec$) with respect to the center of the map in the $-$5.4 \kms\, channel and at an offset ($-4\arcsec$,$-2\arcsec$) in the 2.6 \kms\, channel.

\subsection{The clump B2}

In the southern zone of the blue lobe (not mapped in CS and $^{34}$SO), the interferometric observations show the presence of three clumps. The faintest clump is B2c, detected only in HCN (1--0) and CH$_3$OH (2$_{\rm K}$--1$_{\rm K}$). It is located between B1 and B2 and its shape and size are more similar to the northern clumps than the other two B2 sub-clumps. In fact, B2a and B2b show observational characteristics slightly different with respect to the other clumps. They are detected in all the molecules, with very similar intensity. In all transitions B2a and B2b have a line peak velocity ($v$ $\sim$ 2 \kms) closer to the ambient velocity $v_{\rm lsr}$ = 2.7 \kms\, while all other clumps have a typical peak velocity of $\sim$ 0 \kms. The final velocity of the blue line wing is smaller of about 2--4 \kms\, in B2a and B2b than in the other clumps. B2a and B2b are the clumps farthest from the protostar and also their morphology is different: they are not quasi-circular or displaced in an arch-like shape but they have an elongated shape of about 10 $\times$ 20 arcsec$^2$ and are displaced along the outflow axis. All these characteristics suggest that B2a and B2b are associated with an older shock episode and they have been slowed down during their propagation in the molecular cloud. The lower velocity of the B2 clumps might also be caused by projection effect related to the precession of the outflow axis.

\section{Column densities}

We derived the molecular column density in the detected clumps, from the integrated intensity of the observed emission lines, assuming LTE condition and that the lines are optically thin. In this case the following formula can be used
\begin{equation}
N=1.67\times10^{14}\, \frac{Q(T_{\rm rot})}{\mu^2\nu S}\, exp\left(\frac{E_{\rm up}}{kT_{\rm rot}}\right)\int{T_{\rm MB}\, dv}
\end{equation}
where $Q(T_{\rm rot})$ is the partition function, $\mu$ is the dipole moment in debyes, $\nu$ is the frequency in GHz, $S$ is the line strength, $E_{\rm up}$ is the energy of the upper level of the transition and the integral of the line emission is in K \kms. The integration limits used to calculate the column density is the same for all molecules in each clump and it has been determined using the molecule where the clump is best defined. The column densities for the observed molecules are reported in Table \ref{column}. 
These values must be considered as lower limits because:
{\it i)} the LTE assumption may be not valid for most of the transitions since they have a critical density $\ge$10$^5$ cm$^{-3}$; and
{\it ii)} the assumed temperature of 60 K for B2a and B2b and 80 K for the others clumps, derived by \cite{tafalla95} by means of VLA observations of NH$_3$ emission, are lower limits since at these high temperatures the ammonia rotational transitions are probably not thermalized \citep{danby88} and the real gas kinetic temperature may be significantly higher.
Moreover, the HCN column density is significantly underestimated because of the strong self-absorption of the observed line.
For the calculation of the CH$_3$OH column density we used the (2$_1$--1$_1$)A$^-$ line at 97.583 GHz in the clumps where it has been observed (i.e. B0d, B0e, and the B1 sub-clumps). For the other clumps the situation is more difficult because we must use the (2$_{\rm K}$--1$_{\rm K}$) data and the three lines at 96.739, 96.741 and 96.745 GHz are closely blended (see Fig. \ref{spettri}). However, to roughly evaluate the methanol column density in the whole lobe we used the most intense line (2$_0$--1$_0$) A$^+$ at 96.741 GHz. The integrated line intensity for this line has been calculated summing the emission in a velocity range that has been defined by eye in each clump in order to minimize the contribution of the adjacent lines. However, we stress that in most cases this contribution is not negligible, leading to an overestimate of the methanol column density as it can be seen by comparing the values derived using this line with those derived by using the (2$_1$--1$_1$)A$^-$ line in the clumps where it is observed (column 5 vs column 6 of Table \ref{column}): the first values are always higher (never more than a factor of 10) than the second.

\begin{table*}
\caption{Molecular column densities in the clumps.}
\vspace{0.2cm}
\begin{tabular}{@{}lccccccc@{}}
\hline
clump   & N(HC$_3$N) & N(HCN) & N(OCS) & N(CH$_3$OH)$^{\dag}$& N(CH$_3$OH)$^{\ddag}$ & N($^{34}$SO) & N(CS)\\ 
      & (cm$^{-2}$)  & (cm$^{-2}$) & (cm$^{-2}$) & (cm$^{-2}$)  & (cm$^{-2}$) & (cm$^{-2}$) & (cm$^{-2}$)\\
\hline
B0a    &  2.0$\times$10$^{13}$ & 1.3$\times$10$^{15}$ &...                   & 9.7$\times$10$^{15}$ & ... 		  & ...  		 & ...		       \\
B0b     &  ...		      & ...		     &...                   & 1.7$\times$10$^{15}$ & ... 		  & ...  		 & ...		       \\
B0c     &  ...		      & ...		     &...                   & 9.7$\times$10$^{15}$ & ... 		  & ...  		 & ...		       \\
B0d     & ...		      & 2.4$\times$10$^{13}$ & 6.4$\times$10$^{13}$ & 1.2$\times$10$^{16}$ & 3.6$\times$10$^{15}$ & ...  		 & 9.3$\times$10$^{13}$  \\
B0e    &  1.7$\times$10$^{13}$ & 1.1$\times$10$^{15}$ &...                   & 5.5$\times$10$^{15}$ & 2.3$\times$10$^{15}$ & ...  		 & 5.3$\times$10$^{14}$\\
B1a   &  3.1$\times$10$^{13}$ & 1.8$\times$10$^{15}$ &...                   & 3.5$\times$10$^{16}$ & 3.4$\times$10$^{15}$ & 3.6$\times$10$^{13}$ & 5.5$\times$10$^{14}$\\
B1b   &  3.4$\times$10$^{13}$ & 5.0$\times$10$^{14}$ & 1.5$\times$10$^{14}$ & 5.0$\times$10$^{16}$ & 1.2$\times$10$^{16}$ & 8.3$\times$10$^{13}$ & 4.1$\times$10$^{14}$\\
B1c   &  3.4$\times$10$^{13}$ & 1.7$\times$10$^{15}$ &...                   & 4.0$\times$10$^{16}$ & 3.5$\times$10$^{15}$ & 2.6$\times$10$^{13}$ & 6.4$\times$10$^{14}$\\
B1d   &  2.2$\times$10$^{13}$ & 9.3$\times$10$^{14}$ & ...                  & 3.8$\times$10$^{16}$ & 5.0$\times$10$^{15}$ & 9.0$\times$10$^{13}$ & 3.1$\times$10$^{14}$\\
B2c    & ...		      & 6.3$\times$10$^{13}$ &...                   & 9.3$\times$10$^{15}$ & ... 		  & ...  		 & ...\\
B2a   &  1.7$\times$10$^{13}$ & 4.7$\times$10$^{14}$ & 7.9$\times$10$^{13}$ & 1.7$\times$10$^{16}$ & ... 		  & ...  		 & ...\\
B2b   &  9.2$\times$10$^{12}$ & 4.4$\times$10$^{14}$ & 7.1$\times$10$^{13}$ & 1.7$\times$10$^{16}$ & ... 		  & ...  		 & ... \\
\hline
\end{tabular}
\label{column}
\\
$^{\dag}$ derived from the (2$_0$--1$_0$)A$^+$ line\\
$^{\ddag}$ derived from the (2$_1$--1$_1$)A$^-$ line
\end{table*}

The CS and HC$_3$N column densities are quite uniform in all the clumps, (3--6)$\times$10$^{14}$ cm$^{-2}$ and (1--3)$\times$10$^{13}$ cm$^{-2}$, respectively. Note that the CS column density in the B1a clump is the sum of the two velocity components detected in this position. On the other hand, the methanol column density shows a higher value, 1.2$\times$10$^{16}$ cm$^{-2}$, in B1b with respect to the typical value of the other clumps, (2--5)$\times$10$^{15}$ cm$^{-2}$. On the contrary, the HCN column density has one of the lowest values 5$\times$10$^{14}$ cm$^{-2}$ in B1b while it is higher in B1a and B1c, $\sim$ 2$\times$10$^{15}$ cm$^{-2}$.  The $^{34}$SO column density is $\sim$ 3$\times$10$^{13}$ cm$^{-2}$ in B1a and B1c while it is slightly higher, (8--9)$\times$10$^{14}$ cm$^{-2}$, in B1b and B1d. In all the species, the column densities in B2a and B2b have similar values and are lower than the values in B1. Note, however, that the size of the clumps where the column densities are calculated are larger in B2a and B2b than in the other clumps.

The column densities derived by merging our PdBI data with previously obtained IRAM-30m data for HCN and CH$_3$OH are similar (difference of 10\% on average) to the values derived by \cite{bachiller97} in B1 and B2 using only single dish IRAM-30m data. This shows that most of the large scale emission filtered out by the interferometer has been recovered adding the single dish data to the interferometric data.

\section{Discussion}
\label{disussion}

\subsection{The clumpiness of the blue lobe}

The two species for which the single dish data have been added to the interferometric data, namely HCN and CH$_3$OH, show the presence of a diffuse gas component that matches the morphology of the blue lobe observed in other species such as CO, CS, SiO and SO (\citealt{gueth96}; \citealt{bachiller01}).
Our high spatial resolution maps show that the outflow emission is indeed continuous, with a number of local peaks. In fact, the large clumps detected by previous single dish observations have been resolved into several smaller clumps and few new clumps have been detected along the lobe. These findings confirm what suggested by previous studies (\citealt{viti04}; \citealt{benedettini06}) that several gas components are present in the low velocity molecular clumps observed in chemically active outflows with single dish telescopes.

Two main structures are identified (see Fig. \ref{all_lobe}): the southern one is associated with the B2 clumps and the northern one is associated with the B0 and B1 clumps. These two structures of diffuse molecular gas are associated with the two cavities identified in the CO (1--0) map of \cite{gueth96}. The two cavities have probably been created by the propagation of large bow-shocks due to a highly collimated, precessing jet. 
It is worth noting the two continuous structures perpendicular to the flow axis: the first connecting B0a, B0c and B0d and the second slightly to the north at the level of B0b; they seem to be detected in HCN and CH$_3$OH as well as CO. The nature of these structures remains unclear.

\begin{figure*}
\includegraphics[width=12cm,angle=-90]{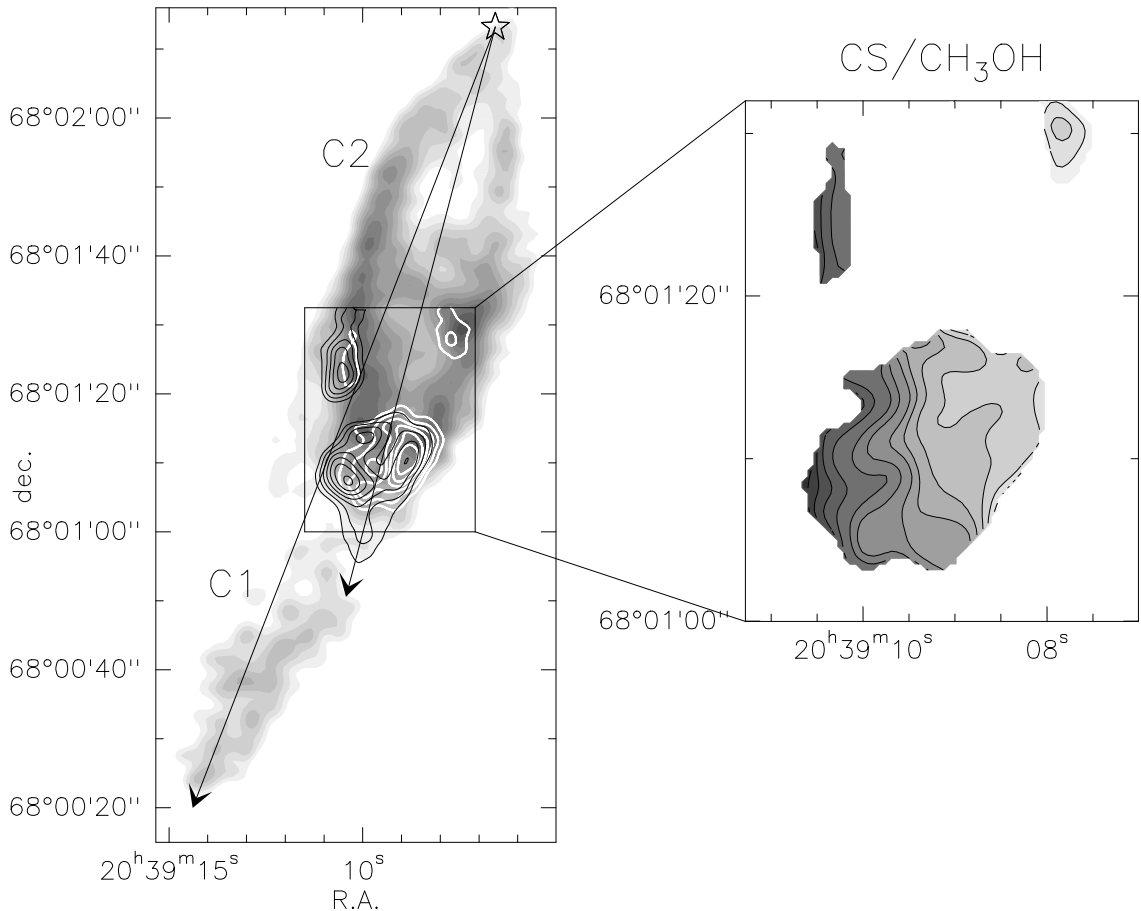}
\caption{Left panel: overlay of CO (1--0) (grey scale) \citep{gueth96}, CS (2--1) (black line) and  CH$_3$OH (2$_1$--1$_1$)A$^-$ (white line). The star marks the position of the protostar and the two straight lines indicate the direction of the precessing jet. Right panel: ratio of the CS and CH$_3$OH (2$_1$--1$_1$)A$^-$ emission; contours are: 10, 20, 30, 40, 50, 60, 80, 100, 150, 200, 250, 300.}
\label{cavity}
\end{figure*}

In Fig. \ref{cavity} we show the CS (2--1) and CH$_3$OH (2$_1$--1$_1$)A$^-$ maps superimposed to the CO (1--0) map. The clumps identified in our maps are located at the walls of the C2 cavity. In particular, B0a and B0e trace the east wall, B0b  and B0d  trace the west wall and the arch-shaped B1a, B1b, B1c and B1d clumps are located at the apex of the cavity. Our maps support the scenario that the cavity was created by the propagation of large bow-shock, showing the presence of clumpy structures at the walls of the cavity with the arch-shaped clumps at the apex that may trace the front of the bow-shock. The situation in the older C1 cavity is less clear since this cavity is less defined in the CO map. However, it seems that the B2a and B2b clumps trace the west wall of the cavity.

\subsection{East/West chemical difference}
\label{disussion_clumps}

The observed species peak in different parts of the C2 cavity. Considering also other interferometric maps found in the literature [SiO (2--1), spatial resolution $\sim$ 2.5 arcsec \citep{gueth98} and NH$_3$ (3,3), (1,1), spatial resolution $\sim$ 5 arcsec \citep{tafalla95}], we can identify two groups of molecules, one peaking in the west-side and the other peaking in the east-side. In particular, the east clumps B0a, B0e, B1a and B1c are brighter in HC$_3$N (11--10), HCN (1--0), CS (2--1) and NH$_3$ (3,3) and SiO (2--1) while the west clumps B0b, B0d  and B1b are brighter in CH$_3$OH (2$_{\rm K}$--1$_{\rm K}$), OCS (7--6) and $^{34}$SO (3$_2$--2$_1$). This peak displacement in the line emission suggests an inhomogeneity in the physical and chemical conditions of the observed clumps. 

The different behaviour between the east and west parts of the map is
also clear from the column densities (see Table \ref{column}). N(CS) and N(HC$_3$N) are quite constant in all the clumps; N(HCN) is higher in the east clumps while N(CH$_3$OH), N(OCS) and N($^{34}$SO) are higher in the west clumps. In particular, the OCS and CH$_3$OH column density shows the highest value in B1b while here the HCN column density assumes its lowest value. The $^{34}$SO column density is slightly higher in B1b and B1d than in B1a and B1c. Despite the large uncertainties associated with the derived column densities, these differences suggest that a chemical differentiation is present among the clumps associated with the C2 cavity. In particular, the N(CH$_3$OH)/N(CS) ratio is higher in the west part (38 in B0d, 29 in B1b and 16 in B1d) than in the east part (6 in B1a and B1c and 4 in B0e) (see Fig. \ref{cavity}). Similarly N($^{34}$SO)/N(CS) is higher in the west (0.20 in B1b and 0.29 in B1d) than in the east (0.06 in B1a and 0.04 in B1c). Since the CS column density is quite constant throughout the map, the difference in the ratio is an indication of an higher chemical abundance of methanol and sulfur oxide in the west part of C2 cavity, especially in B1b for CH$_3$OH and in B1d for $^{34}$SO. This result is independent on the assumption that the temperature of 80 K is constant throughout the B1 clump. In fact, 80 K was derived from the non uniform NH$_3$ (3,3)/(1,1) line ratio by \cite{tafalla95}. In particular, the reference value of 80 K is suitable for the east clumps where the ammonia peaks but it may not be suitable for methanol, which peaks in the west. However, by changing the temperature from 10 K to 200 K, the methanol column density changes only by a factor of 3, and the N(CH$_3$OH)/N(CS) ratio is always significantly higher at west than at east.

\begin{figure*}
\includegraphics[width=10cm,angle=-90]{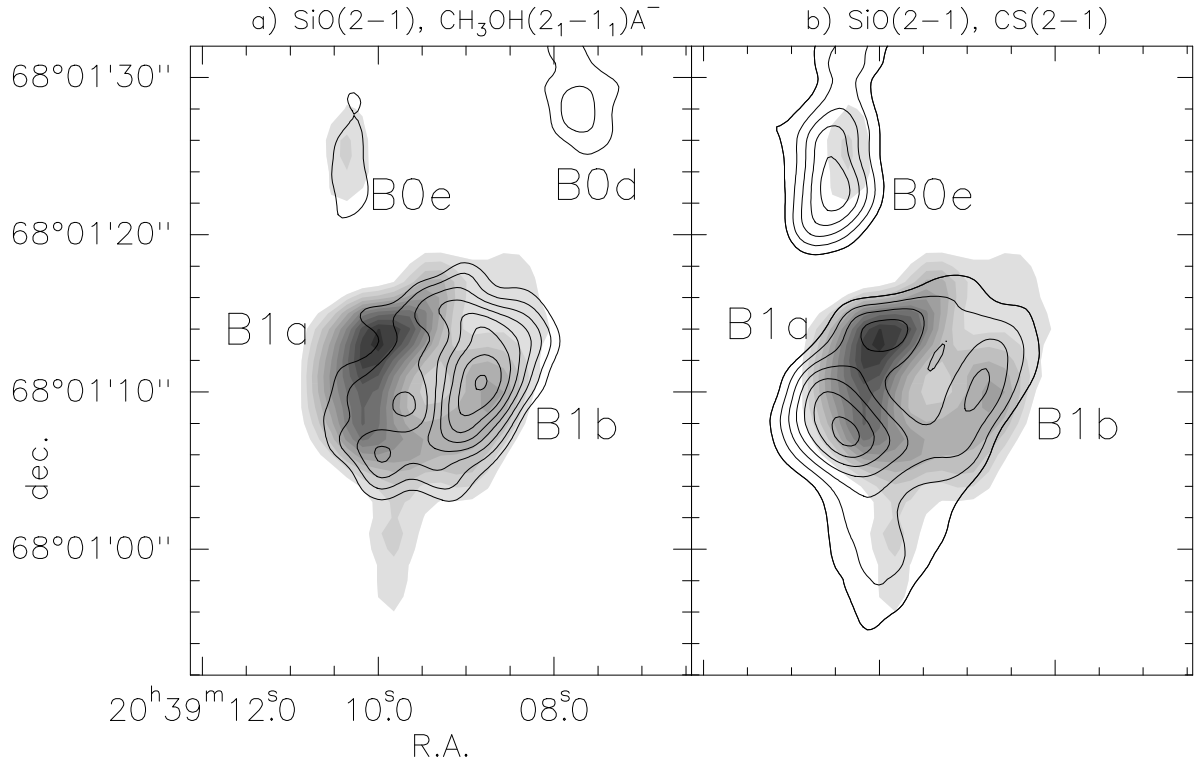}
\caption {Overlay of the SiO (2--1) (grey scale) \citep{gueth98} with: {\bf a)} CH$_3$OH (2$_1$--1$_1$)A$^-$ (continuous line) and {\bf b)} CS(2-1) (continuous line). For SiO the first level is 1 Jy beam$^{-1}$ \kms (3$\sigma$), level steps are 0.5 Jy beam$^{-1}$ \kms;  for CH$_3$OH (2$_1$--1$_1$)A$^-$, first level is 0.01 Jy beam$^{-1}$ \kms (3$\sigma$), level steps are 0.007 Jy beam$^{-1}$ \kms; for CS (2--1), first level is 0.027 Jy beam$^{-1}$ \kms (3$\sigma$), level steps are 0.02 Jy beam$^{-1}$ \kms.}
\label{sio_met}
\end{figure*}

\subsection{The B1 internal structure}

The asymmetry in the line emission is particularly evident in B1. In Fig  \ref{sio_met} we compare the emission of CS (2--1), CH$_3$OH (2$_1$--1$_1$)A$^-$ and SiO (2--1). CS and SiO, as well as NH$_3$ \citep{tafalla95}, show a similar behaviour, peaking $\sim$ 10 arcsec east of the peak of CH$_3$OH. Since the B1 clump is located at the apex of the bow shock produced by the second shock episode, it is particularly interesting to note the different spatial distribution of the SiO, NH$_3$, and CH$_3$OH molecules that are common shock tracers.
SiO (2-1) and CH$_3$OH (2$_{\rm K}$--1$_{\rm K}$) are expected to be excited by the same physical conditions since their critical density and excitation temperature are similar (n$_{\rm crit} \sim$ 10$^6$ cm$^{-3}$ and T$_{\rm ex} \sim$ 4--15 K). It is then probable that a different chemistry is at the origin of the observed displacement, although small differences of the gas density and temperature are also possible between the subclumps.
In fact, methanol and ammonia are abundant on grain ice mantles \citep{charnley92} and they evaporate in the gas phase as the temperature increases above $\sim$ 120 K \citep{collings04}, while SiO is rapidly formed in the gas phase after the sputtering of significant amounts of silicon from the dust grains in C-shock with speeds larger than 25 \kms (\citealt{schilke97}; \citealt{pineau97}). Since the efficiency of the Si sputtering increases with the velocity of the shock, the higher brightness of the SiO emission in the eastern part of the C2 cavity might suggest that the velocity of the shock is higher in the east wall of the cavity with respect to the west wall. This is also supported by the fact that, in general, the terminal velocity of the blue wing for the observed lines is higher, from 4 up to 10 \kms, for the eastern clumps than for the western clumps (see Fig. \ref{spettri}). On the other hand, the chemical models developed for the low-velocity molecular clumps in outflows \citep{viti04} show that the methanol chemical abundance is higher in models with high gas density. In particular, the abundance increases by more than one order of magnitude as the gas density increases from 10$^5$ cm$^{-3}$ to 10$^6$ cm$^{-3}$ (see Table 5 and Fig. 3 of \citealt{viti04}). This suggests that the west clumps might have a slightly higher density, n$_{\rm H_2}$ $\sim$ 10$^6$ cm$^{-3}$, than the mean value n$_{\rm H_2}$ = 3$\times$10$^5$ cm$^{-3}$ derived in B1 from the LVG fitting of the SiO lines observed with the IRAM-30m and JCMT telescopes \citep{nisini07}.

The asymmetry in the emission of the shock tracing species may be explained in the framework of the morphology of the blue lobe of the L1157 outflow. The lobe is composed by two big cavities created by two subsequent bow shocks driven by a precessing jet \citep{gueth96}: the first shock episode is associated with the C1 cavity and the southern clumps and the second shock episode is associated with the C2 cavity and the northern clumps. The model
by \cite{gueth96} found that the jet precesses on a cone with opening angle of 6$\degr$ and that the C1 cavity lies mainly on the plane of the sky while C2 is elongated towards us and towards the west. They also found that between the two shocks the jet precessed a quarter of the rotation period. 
Due to the precession of the jet axis from east to west, the west wall of the C2 cavity is expanding into the denser medium of the harboring molecular cloud while the east wall is expanding in a medium already swept-up by the older shock episode. Therefore chemical differentiations between the east and west clumps are expected because of the different physical conditions of the pre--shock gas. In this scenario one would expect a higher SiO abundance at east, where we observe a higher velocity, and higher CH$_3$OH abundance at west, where the density is likely to be higher. 
On the other hand, the change of the outflow inclination with respect to the plane of the sky between the two shock episodes, even if quite small ($\sim$ 6$\degr$), could imply that the eastern part of B1 may intersect the dense wall of the previously formed C1 cavity. If this is the case, one would expect a higher gas density at the east of B1, i.e. an opposite scenario with respect to the previous one.
A similar puzzling finding is observable in the south part of B1 where a linear structure of diffuse emission is detected in CS (2-1) and in SiO (2-1) but not in CH$_3$OH (2$_1$--1$_1$)A$^-$ (see Fig. \ref{sio_met}). Also in this case CS and SiO trace the same gas component and they are decoupled from the methanol emission. \citet{gueth98} suggested that the SiO finger, that lies along the jet direction, traces the magnetic precursor of the shock. An enhancement of the SiO abundance due to the interaction of the magnetic and/or radiative precursor of the shock with the ambient gas has also been observed in the L1148-mm outflow \citep{jimenez-serra04}.

Finally, it is worth noting that a displacement between the CH$_3$OH (2$_1$--1$_1$) and SiO (2--1) emission has been also observed in the outflow driven by the high--mass protostar IRAS 20126+4104 (\citealt{cesaroni05}, see their Fig. 16). Also in this case, the authors interpret such displacement by the light of the outflow precession with the methanol abundance increased as a consequence of the entrainment of the surrounding high density cloud.
In conclusion, the real nature of the observed displacement of the shock tracer species in B1 can be explained only with a proper determination of the actual physical conditions (density as well as temperature) of the clumps. In particular, to this aim observations of higher excitation CS transitions are needed to constrain possible density variations.

\section{conclusions}

High spatial resolution maps of the blue lobe of the L1157 outflow, obtained with the PdBI have been presented. The blue lobe has been mapped in CH$_3$OH (2$_{\rm K}$--1$_{\rm K}$), HC$_3$N (11--10), HCN (1--0) and OCS (7--6) at a spatial resolution of $\sim$ 6$\times$5 arcsec$^2$. Moreover, the bright B1 clump has also been observed in CS (2--1), CH$_3$OH (2$_1$--1$_1$)A$^-$, and $^{34}$SO (3$_2$--2$_1$) at higher spatial resolution of $\sim$ 3$\times$3 arcsec$^2$.

These observations show a very rich structure in all the tracers, revealing a clumpy structure of the gas, richer than the one observed in previous single dish observations. We detect new clumps, labeled B0b, B0d, and B2c. Moreover, the previously known B0, B1 and B2 clumps have been resolved into three (B0a, B0c, B0e), four (B1a, B1b, B1c and B1d) and two (B2a and B2b) sub-clumps, respectively. These results confirm that the shocked molecular gas in chemically active outflows is mainly structured in small clumps with size of the order of 0.02 -- 0.04 pc.

The two southern clumps (B2a and B2b) of the blue lobe of the L1157 outflow appear similar both in the line intensity and the elongated shape (20$\times$10 arcsec$^2$). They show observational properties slightly different with respect to the other clumps, probably due to their older age. The northern clumps (B0a, B0b, B0c, B0d, B0e, B1a, B1b, B1c and B1d) have typical size of 10 -- 15 arcsec and they are located at the walls of the C2 cavity formed by the propagation of a bow shock, with the B1 sub-clumps tracing the apex of the bow shock and the B0 sub-clumps located in the wake of the shocked region where the matter has already been affected by the passage of the shock. The observed species peak in different parts of the cavity. By taking into consideration other interferometric maps found in the literature, we identify two groups of molecules: HC$_3$N, HCN, CS, NH$_3$ and SiO peak at the west-side and CH$_3$OH, OCS and $^{34}$SO peak at the east-side of the cavity. This peak displacement is also clear from the column densities and suggests an inhomogeneity in the physical conditions and/or chemical composition of the observed clumps. The observed asymmetry may be related to the precession of the outflow: because of the precession of the outflow axis from east toward west, the west wall of the C2 cavity is expanding into the denser medium of the harboring molecular cloud so the velocity of the shock may be lower and the density of the clumps higher than at the east wall which is expanding in a medium already swept-up by the older shock episode. However, other hypotheses cannot be ruled out, therefore the determination of the nature and the physical conditions of the clumps require further high spatial resolution observations and the use of a detailed chemical and shock model.

Finally, we note that the shocked gas in the L1157 outflow will be extensively studied by the PACS and HIFI instruments on board the {\it Herschel} satellite, that will be launched in 2008. In particular, the blue lobe is the chosen position of an unbiased spectral survey in the $\sim$ 450--1990 GHz window with a spatial resolution ranging from $\sim$ 9 to 46 arcsec. Even if the {\it Herschel} observations will be at lower spatial resolution with respect to the interferometric data presented here, the multiline analysis of data from the millimeter to the far-infrared spectral range will allow to trace all the gas excitation from few tens to few thousands of Kelvin and to fully constraint the physical properties of the warm gas components along the chemically active L1157 outflow.

\section*{Acknowledgments}

We acknowledge Riccardo Cesaroni and Paola Caselli for helpful discussions. This work is based on observations carried out with the IRAM Plateau de Bure Interferometer. IRAM is supported by INSU/CNRS (France), MPG (Germany) and IGN (Spain). For the data reduction MB has benefited from research funding from the European Community's Sixth Framework Programme. SV acknowledges financial support from an individual PPARC Advanced Fellowship.

\label{lastpage}

\end{document}